# Power aspects of processes at the piston shock region

P.A. Sedykh

Siberian Branch of Russian Academy of Sciences, Institute of Solar-Terrestrial Physics, Lermontov str., 126 a, p/o box 291, Irkutsk, 664033, Russia

E-mail: pvlsd@iszf.irk.ru

**Abstract**

Identifying blast-wave shocks, which can arise during CME formation, is a much more complex problem. The difference from piston shocks is that a blast-wave shock originates from the explosions that frequently accompany CME formation, and further propagates freely without any CME piston effects. Earth's bow shock (BS) is a piston shock. Behind the bow shock front there is a flow of the modified solar wind plasma: transition layer, which also carries the modified magnetic field of solar wind. Velocity and density of plasma as well as parameters of magnetic field of this current can be estimated if the form of the bow shock front and of magnetopause are considered to be known. In this paper we assumed them to be paraboloids of rotation. In this paper we have determined potential distribution along magnetopause from the balance condition of substance coming into transition layer from the solar wind on one side and leaving through the gap between magnetosphere and the bow shock front and through magnetopause on another. To a first approximation this distribution differs from potential distribution at the BS front only in a constant multiplier. We used the established potential distribution as a boundary condition while solving the problem on potential distribution in the magnetosphere. The first solution harmonic turned out to coincide in the form with the boundary condition on magnetopause. We have constructed the full solution for electric field in the magnetosphere having limited ourselves to the first harmonic and corotation field. We have obtained necessary equations to model processes in the region of bow shock.

## 1. Introduction

Scientific and technical research into shocks, forming in outer space and on the Earth, may be very useful. Shock-wave research was started as far back as the middle of the XIXth century by a group of mathematicians dealing with applied problems and by a group of experimenters. Over the period between the Second World War and until the end of the 20th century, there arose a need for new and accurate data on blast waves from powerful explosions and shocks generated by spacecraft entering dense atmospheric layers and by supersonic flights. The need gave a new impetus for many studies by famous scientists. Vast literature which contains a considerable amount of information, collected in this field of science for many decades, has been published.

A blast-wave shock occurs when in a substance (plasma, gas, liquid, and solid) energy is suddenly released or injected into it from outside, triggering an explosion. For simple explanation, we will consider a spherical explosion. During a spherical explosion, the process proceeds as follows. When energy is suddenly released in the air, water, underground or in plasma, a spherical volume of pressurized hot gas is formed which, when expanding, causes a shock in the environment. The shock moves like a sea tidal wave and immediately raises pressure, density, and temperature of the environment. It also produces a substance flow that follows the shock. Covering ever larger amounts of substance, the shock gradually decays into a weak disturbance or a sound wave. The decay occurs very quickly nearby the explosion center when the shock is very strong and pressure, temperature, density, and velocity behind it are extremely high, and almost ceases when the shock turns into a sound pulse. The sphere, heated by the explosion, is expanding and pulsing until it reaches an equilibrium size corresponding to the environmental pressure. Then the heated substance in the environment diffuses.

Our planet can be represented as a piston moving through the solar wind (see Fig.1). Only in this case, we deal with plasma; at the same time we should account for the presence of the interaction between the Earth's magnetic field and the interplanetary magnetic field (frozen-in to the solar wind plasma). The processes occurring in the bow shock region will be even more complex. Earth's bow shock (BS) is a piston shock.

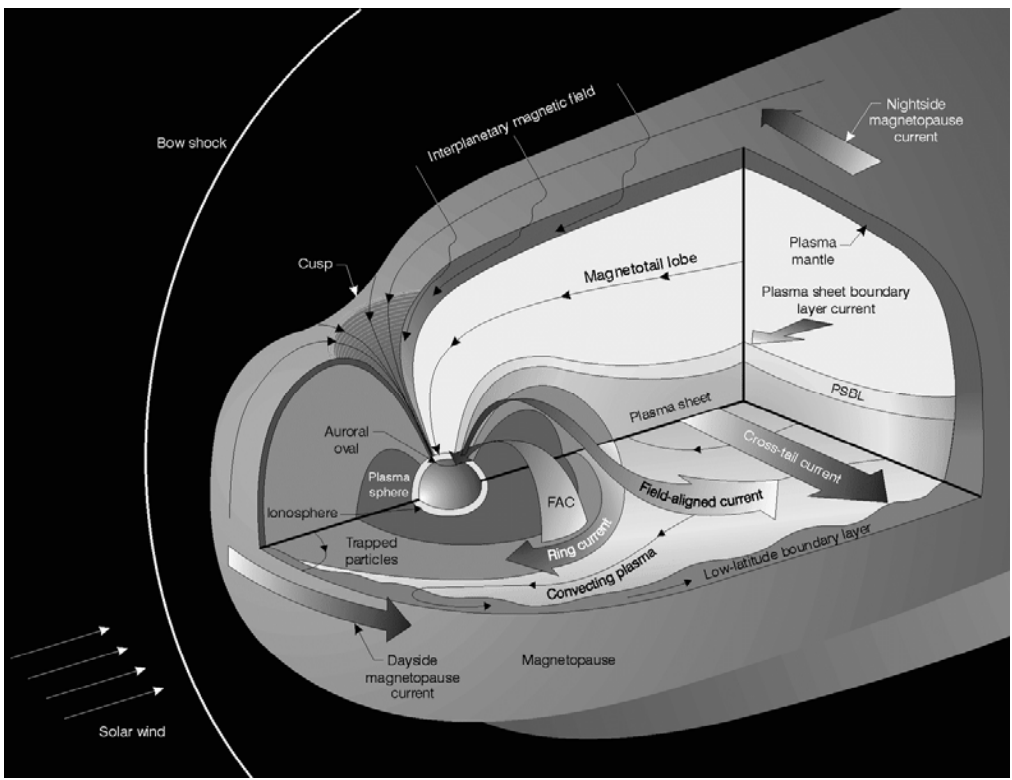

**Figure 1.** The front of Earth's bow shock and the structure of the geomagnetosphere (Kivelson and Russell, 1995).

Features of blast-wave and piston shocks are in detail stated in (M.V. Eselevich, Eselevich V.G., 2011). A method for detecting and identifying piston shocks preceding coronal mass ejections (CMEs) and propagating in various directions relative to the ejections is presented in (M.V. Eselevich, Eselevich V.G., 2011). The physics behind this method is based on the concept of a perturbed zone (M.V. Eselevich, Eselevich V.G., 2011) that arises in front of the CME, due to its interaction with the background solar wind. When the CME velocity relative to the velocity of the unperturbed solar wind exceeds the local magnetoacoustic speed, which is approximately equal to the Alfven speed in the solar corona, a piston shock forms in the leading portion of the perturbed zone. In this case, the CME is the piston. Identifying blast-wave shocks, which can also arise during CME formation, is a much more complex problem. The difference from piston shocks is that a blast-wave shock originates from the explosions that frequently accompany CME formation, and further propagates freely without any CME piston effects.

Since early 1960s many experimental and theoretical investigations of the supersonic and super-Alfvenic solar wind flow past the Earth considered the highly dynamic bow shock boundary upstream of the planetary magnetopause ((Beard, 1960); (Spreiter, et. al., 1966); (Fairfield, 1971); (Cairns, 1995); (Brecht, 1997)). A relevant review of pertinent studies can be found in the paper by Peredo et al. (1995) presenting the empirical model of the Earth's bow shock.

The purpose of this paper is to determine necessary equations for modeling processes in the region of piston shocks. I will use partially the results from (Sedykh, Ponomarev, 2012), (Sedykh, 2014).

## 2. Important relations and basic theoretical analysis

The basic flow of solar wind energy is a flow of kinetic energy of supersonic plasma current carrying trapped magnetic field. Everywhere, except for the BS front, this plasma can be considered ideal. Inside the front, plasma is extremely strongly turbulized and heated up to the temperature corresponding to average kinetic energy of progressive motion. In the nose point density of plasma and tangential component of magnetic field behind the front are higher in $\eta = (\gamma+1)/(\gamma-1)$ times, than before the front. Here γ is an adiabatic index, $Bt_1/Bt_0 = \sigma$ and is associated with η formula (Korobeinikov, 1985; Whang, 1987):

$$\sigma = \eta(A_0^2 - 1)/(A_0^2 - \eta)$$

where $A_0^2 = 4\pi\rho_0 V_0^2/B_0^2$, that is the relation of dynamic pressure to magnetic in SW, is much greater than 1. Hereinafter the subscripts 0,1,2 will denote the belonging of the given value to solar wind, transition layer(magnetosheath) or magnetosphere, respectively. The difference of tangential component before and behind the front means that electric current flows in the front, and the fact that plasma flow carries solar wind trapped field through the front means that there is electric field in the BS coordinate system. Let us introduce some additional concepts and restrictions. We shall consider the BS front and magnetopause to be confocal paraboloids of rotation. In fact, the BS front is a triaxial hyperboloid, and magnetopause has an even more complex form, but their difference from a paraboloid is not too large and does not have a basic character. Of principal importance is, that both the form of the bow shock and magnetopause should be determined from conditions of interaction of solar wind with the Earth magnetic field. But this is another problem, and for the sake of simplicity and clearness we have decided to sacrifice self-consistency when solving the basic problem. We shall use the solar-magnetospheric coordinate system, whose axis X is directed towards the Sun, axis Y from dawn to dusk, plane XY coincides with the plane of the ecliptic (Fig. 2). Axis Z complements the coordinate system and coincides with the axis of terrestrial magnetic dipole. We represent geomagnetic field in the form, suggested by Antonova and Shabansky (1968), but without taking into account tail current. We shall represent geomagnetic field in the spherical coordinate system r, θ, λ. Besides, we shall use two more auxiliary coordinate systems (Fig. 2). Let some plane pass through axis X and make angle ψ to axis Z. Vector **r**, starting with the beginning of the coordinate system and marking the given point by its end, lays in this plane.

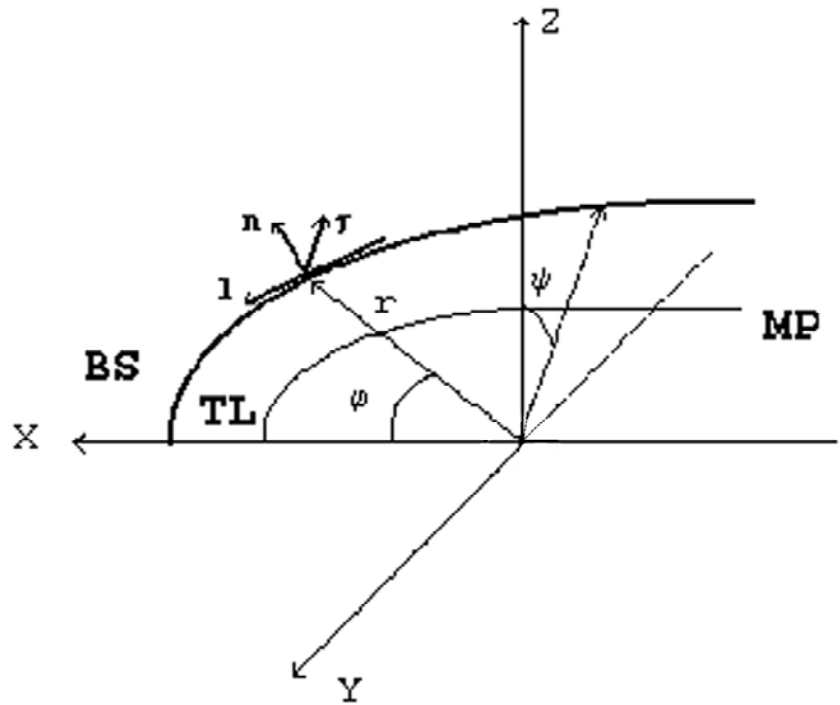

**Figure 2.** A sketch of the Bow Shock (BS), of the Transition Layer (TL) or magnetosheath, magnetopause (MP). The location of the orthogonal coordinate system (axis x is directed towards the sun, axis y is directed from dawn to dusk and axis z is directed towards the geographic north) and the location of auxiliary coordinate system (r, φ, ψ) and the local coordinate system (l, n, τ).

Let us denote the angle between this vector and axis X by φ. It will be the third coordinate of our auxiliary system. Let us introduce the local coordinate system. Its axis l will coincide with the tangent at some point of paraboloid. Its positive direction is the same as of axis X. The second coordinate n is directed on the normal at the same point, and the third one τ supplements the local coordinate system to the right one. We shall consider magnetic field lines to be equipotential. Then, as shown in (Ponomarev, 1985), plasma tube (i.e. plasma content of the magnetic flux tube) will pass from one flux tube in another without surplus and deficiency during electric drift. This allows to replace examination of plasma tubes' drift in the whole magnetosphere by examination of the drift of equatorial traces of these tubes, that is to reduce three-dimensional problem to two-dimensional one.

Let origins of all coordinate systems (certainly, except for the local) coincide with the origin of the solar-magnetospheric one and let the angle between the tangent to the BS front and axis X be $\alpha$. Then, we can write the components of interplanetary magnetic field in the local coordinate system as (Korobeinikov, 1985; Whang, 1987):

$$B_{l0} = B_0[b_{x0}\sin(\varphi/2) + (b_{y0}\sin\psi - b_{z0}\cos\psi)\cos(\varphi/2)] \qquad (1)$$

$$B_{n0} = B_0[b_{x0}\cos(\varphi/2) - (b_{y0}x\sin\psi - b_{z0}\cos\psi)\sin(\varphi/2)] \qquad (2)$$

$$B_{\tau 0} = B_0[b_{y0}\cos\psi + b_{z0}\sin\psi] \qquad (3)$$

And current density under the arch of the BS front will be:

$$j_{l1} = -c(\sigma-1)B_{\tau 0}/4\pi d \qquad (4)$$

$$j_{\tau 1} = c(\sigma-1)B_{l0}/4\pi d \qquad (5)$$

where d is thickness of the BS front; $\sigma$ is the ratio between magnetic field tangential components in front of and behind the BS front (Fig.3). The corresponding surface current density will be:

$$J_{l1} = -c(\sigma-1)B_{\tau 0}/4\pi \qquad (6)$$

$$J_{\tau 1} = c(\sigma-1)B_{l0}/4\pi \qquad (7)$$

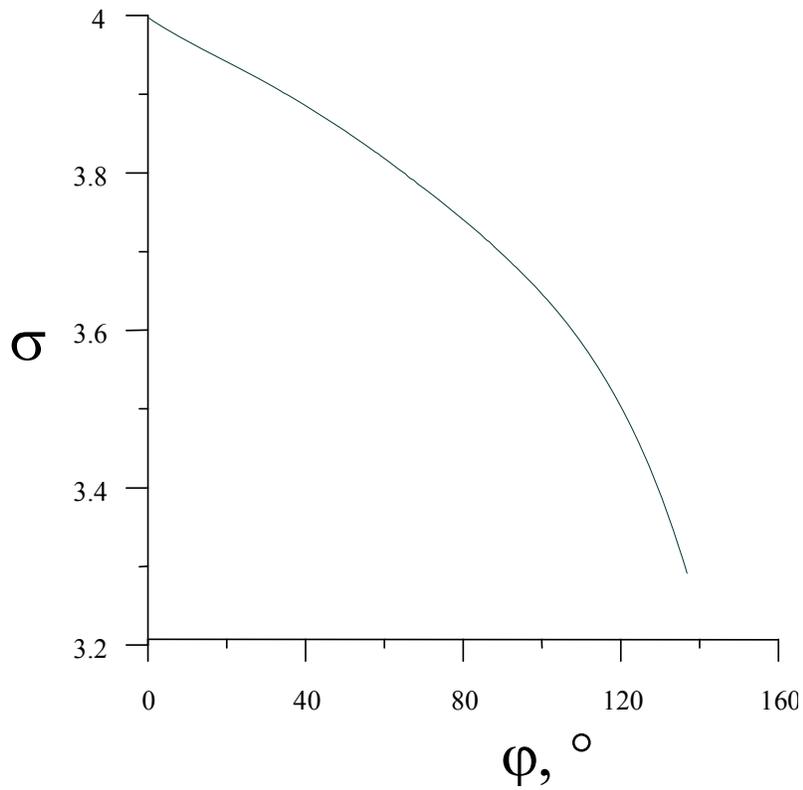

**Figure 3.** Value of transition coefficient $\sigma=B_{lt}/B_{0t}$ as a function of the angle $\varphi$ for the equatorial section (calculated values) (Sedykh, 2014).

These currents' density varies along the front and, therefore, these currents can be divergent. Then there appears a normal current component:

$$j_{n1}= (1/y_g)[\partial(J_{l1}y)/\partial l + \partial J_{\tau 1}/\partial \psi] \qquad (8)$$

where $y_g = r_g \cdot \cos\varphi$, y is coordinate of the BS front point $(r_g,\varphi)$ in the auxiliary coordinate system. Index g denotes belonging of the given value to the BS front. In this system the equation of the front looks especially simple:

$$r = y_g / 2\cos^2(\varphi/2) \qquad (9)$$

where $y_g$ is the distance from the origin of coordinates to the front in the plane YZ, which is equal to the double distance to nose point for the parabola $2x_g$.

In the auxiliary coordinate system the expression for a normal current component will become:

$$j_{n1} = -c[B_{\tau 0}\cos^3(\varphi/2)\partial\sigma/\partial\varphi]/2\pi y_g \qquad (10)$$

Destiny of this current is subject to examination. In principle, any current in a stationary system must be closed. Only two ways of this current closing can be seen: through magnetosphere body or inside the transition layer. Most likely, they both exist simultaneously. If the current is not used in magnetosphere, it closes in the transition layer, additionally accelerating plasma there. Anyway, this current is ready to work on plasma. Let us notice, by the way, that Poynting vector **S** is closely connected with electric current density by the ratio:

$$\mathbf{S} = \mathbf{j}\Phi - (c/4\pi) \text{ curl } (\mathbf{B}\Phi) \qquad (11)$$

Φ – an electric potential.

Now let's turn to determination of electric field at the BS front. Tangential component of this field is:

$$E_{l1} \cong E_{l0} = -V_{n0}B_{\tau 0}/c = -(V_0 B_{\tau 0}/c)\cdot\sin\alpha = -(V_0 B_{\tau 0}/c)\cos(\varphi/2) \qquad (12)$$

Integrating $E_l$ along the front on coordinate l we shall obtain:

$$\Psi_g = \Psi_{g0} \cdot tg(\varphi/2) \tag{13}$$

where $\Psi_{g0} = (V_0 B_{\tau 0}/c) \cdot y_g$.

As one would expect, the electric field during its movement along the front ($\varphi \to \pi$) tends to its undisturbed value, with potential tending to infinity. However, the front itself does not exist far from the disturbing body. Therefore, potential divergence at $\varphi \to \pi$ should not be taken into account. In reality, we shall simply restrict ourselves to the critical angle of 135°, outside of which we shall consider BS not existing. Now let us estimate magnetopause potential. We shall proceed from the following reasons. After solar wind plasma has crossed the bow shock front, it divides into two flows. One is enclosed in the gap between the BS front and magnetosphere, the second crosses magnetopause and sinks into magnetosphere. It should be noted, that there existed an opinion among geophysicists, that magnetopause is impermeable for plasma. It is easy to show, that the density of electromagnetic energy flow is proportional to the mass flow density. Indeed, having substituted the expression for **E** into the expression for **S**, we receive $\mathbf{S} = 2\mathbf{V}p_B$, where $p_B$ is magnetic pressure. On the other hand, mass flow density $\mathbf{I} = \rho \mathbf{V}$, hence $\mathbf{S} = 2\mathbf{I}(p_B/\rho)$. Now we can write down the equation of plasma balance:

$$\int n_0 V_{n0} d\Sigma_g = \int n_1 V_{n2} d\Sigma_m + \int n_1 V_{1l} d\Sigma_b \tag{14}$$

Here:

$$d\Sigma_g = 2\pi y_g \cdot tg(\varphi/2) dl_g \tag{15}$$

$$d\Sigma_m = 2\pi y_m \cdot tg(\varphi/2) dl_m \tag{16}$$

$$d\Sigma_b = \pi(y_g + y_m) \cdot tg(\varphi/2) dh \tag{17}$$

Where, $d\Sigma$ are elements of the BS front surface, magnetopause and gap area, orthogonal to the flow. We have denoted the width element of the gap through $dh$.

Correspondingly:

$$n_0 V_{n0} = (cn_0 B_{\tau 0}/B_0^2) \cdot \partial \Psi_g /\partial l_g, \qquad (18)$$

$$n_1 V_{n2} = (\sigma cn_0 /B_\theta) \cdot \partial \Psi_m /\partial l_m, \qquad (19)$$

$$n_1 V_{11} = -(cn_0 B_{\tau 0}/B_0^2) \cdot [\partial \Psi_g /\partial h - \partial \Psi_m /\partial h] \qquad (20)$$

Substituting (15) - (17) and (18) - (20) into (14) we shall obtain the equation connecting $\Psi_g$ and $\Psi_m$:

$$\Psi_m = a \int [(\partial \Psi_g /\partial \varphi) \cdot d\varphi]/b \qquad (21)$$

where $a = [1 + 2y_g/(y_g + y_m)]$, $b = [1 + 2y_m \sigma B_0^2/(y_g + y_m) B_{\tau 0} B_\theta]$. Here $B_0^2 = B_{10}^2 + B_{\tau 0}^2$, and $B_\theta$ is magnetic field on magnetopause. In reality b varies within small limits and it can be replaced by average value. Then we can obtain:

$$\Psi_m = (a/b) \cdot \Psi_g \qquad (22)$$

Now we can use $\Psi_m (\varphi)$ as a boundary condition for determination of potential inside the magnetosphere. Since, according to our agreement, magnetopause has the form of a paraboloid of rotation, and magnetic flux tubes are equipotential, then as it was mentioned above, it is enough to receive potential distribution in the equatorial plane. Let us introduce the parabolic coordinate system s,t with the centre in the coordinate origin of the solar-ecliptic one. The equation for potential can be written down in a simple form:

$$\partial^2\Psi/\partial s^2 + \partial^2\Psi/\partial t^2 = 0 \qquad (23)$$

Any analytical function of the complex number $z = s + it$ will be the solution (23).

Let's represent the solution as power series:

$$\Psi = \sum[A_n(kz^2)^n + B_n(kz^2)^{-n}] \qquad (24)$$

We can go over to the Cartesian coordinate system using formulas:

$$x = st, \; y = (t^2 - s^2)/2$$

It is convenient to present the argument $kz^2$ in the trigonometrical form: $kz^2 = kr \cdot \exp(i(\varphi+\pi/2))$.

Using function orthogonality, we can find expressions for coefficients:

$$A_n = (1/\pi)\int \Psi_m^0 \cdot tg(\varphi/2) \cdot (kr)^{-n} \cdot \exp(-i[n(\varphi+\pi/2)])d\varphi \qquad (25)$$

$$B_n = (1/\pi)\int \Psi_m^0 \cdot tg(\varphi/2) \cdot (kr)^n \cdot \exp(i[n(\varphi+\pi/2)])d\varphi \qquad (26)$$

Taking into account, that $r = y_m/2\cos^2(\varphi/2)$ on magnetopause, we can integrate these expressions (from 0 to $\pi$). Since integrals (26) diverge, let us restrict ourselves to determination of real U and imaginary I parts of coefficient A. Further we shall restrict ourselves only to the first decomposition harmonic as it is usually the case while analyzing magnetospheric electric field:

$$\Psi_1 = (2/\pi)\Psi_m^0 (r/y_m)U_1\sin\varphi = (2/\pi)\Psi_m^0 (y/y_m)U_1 \qquad (27)$$

It follows from (27), that equipotentials can be lines y = const, that is, lines parallel to axis X. Since magnetospheric electric field is formed by convection and corotation fields, corotation potential should be added to (27):

$$\Psi_c = -\Omega_l R_0^3 B_0/cr \qquad (28)$$

here $\Omega_l$ is circular frequency of the Earth rotation, $R_0$ is radius of the Earth, $B_0$ is intensity of geomagnetic field at the pole. So, total potential is:

$$\Psi = (2/\pi)\Psi_m^0 (r/y_m) U_1 \sin\varphi - \Omega_l R_0^3 B_0/cr \qquad (29)$$

## 3. Conclusion

Let us make estimation of the power separated in the magnetosphere provided that current density $j_{n1}=j_{n2}$, that is, that the whole current generated in the BS penetrates into the magnetosphere. Then having differentiated (27) on y we find $E_y$ and having multiplied by current density from (10), we shall find (for the dawn-dusk meridian):

$$j_{n2} E_y = \cos(\pi/4)\cdot(a/b)\cdot(\partial\sigma/\partial\varphi)\cdot(V_{n0} B_{\tau 0}^2/2\pi y_m) \qquad (30)$$

Having substituted the same values that were used earlier and having taken $\partial\sigma/\partial\varphi$ to be of the order 1 we find that the power separated in the unit of volume of magnetosphere is of the order $3\times 10^{-12}$ erg /cm·s. If we assume power flux to be the same everywhere, and magnetosphere volume is of the order of $10^{30}$ cm$^3$, we get full power of the order of $3\times 10^{18}$ erg/s.

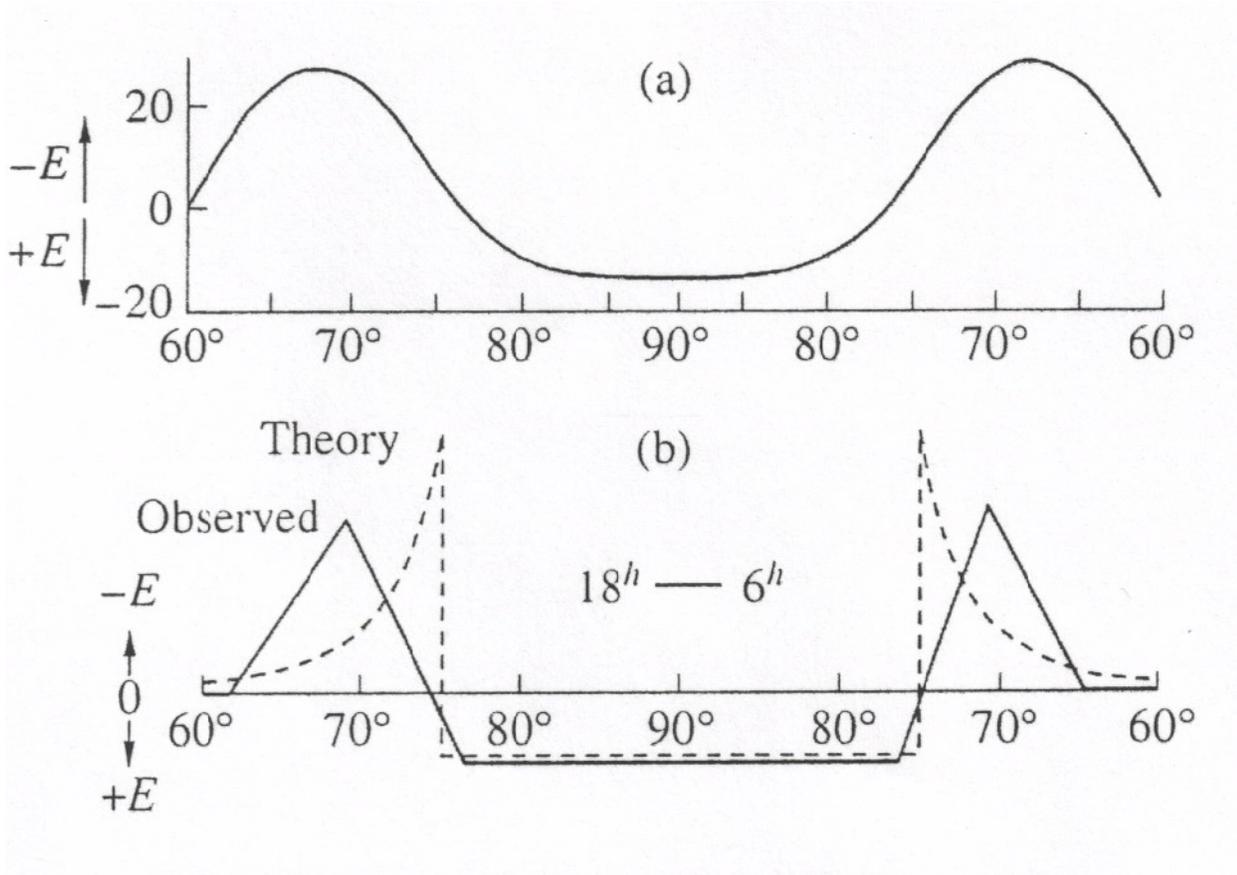

**Figure 4.** The character of the electric field distribution along the dawn-dusk meridian: (a) our calculated values, and (b) data obtained in (Heppner, 1977).

The electric field along the bow shock front and the potential depend on the solar wind velocity, normal component and on the IMF tangential component. The magnetopause potential Ψm is derived from the conditions of balance of the matter coming through the bow shock front and outgoing from the magnetosheath through the magnetopause and the gap between the bow shock front and magnetopause. This potential differs from Ψg only in multiplier. If we assume that the flux tubes are equipotential, the motion of the plasma tube content completely depends on the motion of the tube equatorial trace. Thus, it is sufficient to determine the potential distribution in the equatorial plane within the boundaries, one of which (the magnetopause) is represented by a parabola with the parameter $y_m$ and the other, plasmapause, is determined by a circle of radius $r_p$. The problem can be solved in parabolic coordinates, where the Laplace operator seems to be the simplest. The solution is sought in the form of expansion into the series in terms of orthogonal functions in a standard way. The obtained result is also standard. The character of the electric field distribution over the dawn-dusk meridian fully corresponds to the classical distribution obtained in (Heppner, 1977) (Fig. 4). The power source for maintaining convection was specified, and the boundary conditions at the magnetopause were obtained from solving the general problem rather than they were set from intuitive considerations. The problem to determine the power coming into the magnetosphere is solved as if automatically because **j** and **E** are known. We should merely integrate the product of these values over the volume of the magnetospheric cavity.

Let us imagine the geomagnetic field. The magnetic field lines as if penetrate through the surface of the Earth, within which the field source is located. Let us place the intersection point into an elementary circle. A bundle of field lines within this circle will be called a flux tube. The widest part of this tube falls on the magnetic equator, and the trace of the intersection with the equatorial plane is called the equatorial trace. In the course of the electric drift the plasma tube moves from one flux tube to another tube without excess and deficiency, if magnetic field lines are equipotential. Let us imagine for definiteness that electric drift proceeds toward increasing magnetic field. In such a case, the plasma tube volume constantly decreases, and this means that pressure increases at a uniform adiabatic compression. The electric current from the generator at the bow shock front is responsible for the power needed to compress plasma tubes during their motion toward the Earth.

It is necessary to note, that interest to research processes in interplanetary shocks and in the bow shock has strongly increased recently (researches within the framework of projects GEOTAIL, CLUSTER-II, THEMIS, SPECTR-R). Selecting and applying of correct initial system of equations are very important for modelling processes in the region of piston shocks. In conclusion, let us remind once again, that we have obtained the necessary equations to model processes in the region of bow shock.